\documentstyle[11pt,psfig]{article}

\newcommand\bg{\begin{eqnarray}}
\newcommand\ed{\end{eqnarray}}
\def\ra{\rightarrow}

\def\hs{\hspace{0.1in}}
\def\e{\eta}
\def\be{\bar{\eta}}
\def\D{\partial}
\def\no{\nonumber}
\def\slashD{{/\kern-0.7em\partial}}

\begin{document}

\begin{titlepage}

\title{\bf{
New Numerical Method for \\ Fermion Field Theory}}
\author{
John W. Lawson {\dag}
{\thanks{Present Address:~ICTP, 34014 Trieste, Italy}}
and G.S. Guralnik
{\thanks{Research supported in part by DOE Grant DE-FG02-91ER40688 - Task D}}
\\
Department of Physics \\
Brown University \\
Providence RI 02912}

\date{}
\maketitle
\begin{abstract}
\noindent
A new deterministic, numerical method to solve fermion field
theories is presented.
This approach is based on finding solutions $Z[J]$ to the lattice
functional equations for field theories in the presence of
an external source $J$.
Using Grassmann polynomial expansions for the generating functional $Z$,
we calculate propagators for systems of interacting fermions.
These calculations
are straightforward to perform and
are executed rapidly compared
to Monte Carlo.  The bulk of the computation involves a single
matrix inversion.
Because it is not based on a statistical technique,
it does not have many of the difficulties often encountered
when simulating fermions.
Since no determinant is ever calculated, solutions to problems
with dynamical fermions are handled more easily.
This approach is very flexible, and can be taylored
to specific problems based on convenience and
computational constraints.
We present simple examples to illustrate the method;
more general schemes are desirable for more complicated
systems.

\end{abstract}
\thispagestyle{empty}

\vskip-20cm
\noindent
\phantom{bla}
\hfill{{\bf{BROWN-HET-954}}} \\
\end{titlepage}

\section{Introduction}

     The need for new calculational probes of quantum field theory
is well-known.  Despite the success of various iterative methods including
perturbation theory,
many problems, especially for
strongly coupled systems, persist in both high energy and condensed
matter physics.
The advent of high performance computers has made numerical
attacks on these problems practical.  Most of these
approaches have roots in Monte Carlo methods.  They form the basis
of many calculations in statistical mechanics, including nonrelativistic
field theory, and have been co-opted by high energy physics for
lattice gauge theory calculations \cite{reb1}.

     In high energy physics, Monte Carlo methods had remarkable
initial successes in calculating various quantities associated
with quantum field theory.  Unfortunately, even for simple systems,
calculations involving dynamical fermions,
or performed on large lattices,
become expensive in both time and computer resources.  Despite
refined algorithms and enormous increases in computer power,
progress with Monte Carlo based methods has been slow compared to the
initial impressive successes.  For lattice QCD \cite{reb2},
detailed studies of systems
with dynamical quarks with low masses may require
computer time well beyond current capabilities.

In condensed matter physics,
understanding models of strongly correlated electrons
using numerical methods has serious limitations.
Much of this work has been based on
exact diagonalization
or quantum Monte Carlo.
Diagonalization is powerful especially for calculating dynamical properties,
but in practice, is limited to small systems.  Typically,
4x4 clusters are the largest that can be studied.
Alternatively, Monte Carlo methods can examine larger
lattices, but for
fermionic systems,
simulations face serious problems.
This includes the so-called ``minus sign" problem which makes
defining a positive-definite probability measure
problematic \cite{negele}.
This difficulty plagues
many models, including many of those
relevant to high $T_{c}$ superconductivity.

     We have formulated a numerical scheme for lattice quantum
field theory which does not rely on any statistical method
and is not restricted to small systems
\cite{galerkin,boson,garcia,john}.
This approach is based on considering QFT in the presence of an
external source.
Using either the operator equations of motion, or equivalently the path
integral representation, equations can be derived for the vacuum persistence
amplitude $Z$.  On a lattice, the result is a
set of coupled {\em linear} differential
equations for $Z$ in the discretized sources.
For fermionic problems, these sources will be
anticommuting Grassmann variables.
Once we have obtained $Z$, the lattice Green's functions
can be extracted by functional differentiation.

Despite the large number of coupled differential
equations for $Z$, one for each
source variable, there is nothing unusual about this problem
mathematically.
We can solve them as we do any set of linear differential equations.
Since at the end of any calculation the sources are set to zero,
the obvious first choice is to expand $Z$ in a power series in
the sources about the origin.
For fermionic systems with only a few sites, exact solutions
for $Z$ are possible.
This is due to the self-terminating nature of
Grassmann polynomials.
In general, we cannot solve the theory exactly, and therefore,
we must introduce an approximation.
For the purposes of this paper, we
truncate the series expansion for $Z$
after some maximum power in the source variables.
In this way, we obtain a closed, finite set of linear equations
for the expansion coefficients accurate
to within the limits of the approximation.

The coupled differential equations have a high degree of
symmetry due to the translation and
reflection invariance of the underlying
periodic lattice.  This structure can be used to great advantage
in constructing solutions, greatly reducing the number of equations to be
considered.  This reduced set of algebraic
equations for the coefficients of $Z$
is necessarily inconsistent; this being an artifact of the
approximation.  There are many ways to deal with this problem.  One
particularly elegant method is to require that weighted averages of the
truncated equations vanish.  A clever choice of weight functions
speeds the convergence of the truncated equations to the exact solution.
Numerical techniques of this sort
are called Galerkin methods and are well-known \cite{fletcher}.

     Since our expansion and weight functions depend on the external
sources, we call this method ``the Source Galerkin method"
\cite{galerkin}.  It is different
in concept and application from any other technique that we know of for
solving quantum field theory.
As a practical matter, fermions and bosons are treated
in a similar manner.  Consequently, solutions for dynamical
fermion problems may be possible with existing computers.
On the other hand, this method is not without difficulties.
Principally, the power series grows quickly,
making examination
of large systems difficult.
In these cases, different basis functions must be used.
The examples presented
here illustrate one of the simpler formulations, but not generally
the most useful.

This paper is organized as follows.
Section Two gives a general overview
of the bosonic formulation of the Source Galerkin method.
Section Three outlines the Source Galerkin formulation
for fermions.
The functional formulation is identical
to the boson case except that the sources
become anticommuting Grassmann variables.
The partition function $Z$ satisfies a set
of coupled Grassmann differential equations.
As with bosons, we solve these equations by power series
in the source variables.
In Section Four, we examine
small lattice systems where exact solutions
are possible.
For more realistic systems, the series must be
truncated and a Galerkin procedure devised.  We propose a
Galerkin method for fermions in Section Five
and perform calculations on $1D$
self-interacting fermion systems.
We close by illustrating the
flexibility of the Source Galerkin method
with respect to the choice of expansion functions.
We study $2D$ lattice Gross-Neveu
model by examining the leading behavior of $Z$ as $J \ra 0$.
The first term is equivalent to a Gaussian.  We calculate the
behavior of the chiral condensate as an order parameter for chiral
symmetry breaking.  A method to calculate corrections is outlined.

\section{Source Galerkin Method}

In this section, we review the bosonic formulation
of the Source Galerkin method.
We study quantum field theory in the
presence of an external source.
The vacuum persistence amplitude
$Z[J] = _{J} \langle 0|0 \rangle_{J}$
is the generating functional for the Green's functions.  It
is constrained by a functional differential equation.
For a self-interacting scalar field theory,
the dynamics are described by
\bg
L_{J}\,(\phi) \; = \;
1/2 \, \hat{\phi} \;  (\Box + M^{2}) \; \hat{\phi} \; +
\; g/4 \, \hat{\phi^{4}} \; + \; J \hat{\phi}.
\ed
Taking vacuum expectation values of the
operator equations of motion
\bg
( \Box + M^{2} ) \langle \hat{\phi} \rangle _{J} \; + \;
g \,  \langle \hat{\phi^{3}} \rangle_{J} \;  =  \;
J \,  _{J} \langle 0  | 0 \rangle_{J}
\ed
and identifying Euclidean expectation values
of fields with functional derivatives
of $Z[J]$
\bg
G(x_{1} \ldots x_{n}) \; = \;
\frac{\delta^{n} Z[J]}{\delta J(x_{1}) \ldots \delta J(x_n)}
\ed
yields the functional relation
\bg
( \Box + M^{2} ) \, \frac{\delta Z}{\delta J(x)}
\; + \; g \, \frac{\delta^{3} Z}{\delta J(x)^{3}}  \;  =  \; J \;Z.
\ed
After solving this equation for $Z$, we can extract all
information about our field theory by functional differentiation.
On a $D$-dimensional Euclidean lattice,
the functional equation
becomes a set of coupled differential equations
\begin{equation}
(2 \, D + M^{2}){\D Z \over \D {J(i)}}
 -  \sum_{nn} {\D Z \over \D {J(i)}}
 +  g {{\D^3 Z  }\over {\D {J(i)^3}}}
 =  J(i) Z
\end{equation}
where the sum is over nearest neighbors. There
is one equation per site.

For finite lattices with $N$ sites,
$Z$ as a multivariate function of the $N$
source variables $J(i)$.  To construct a solution
to the differential equations, we can expand $Z$
on any complete set of functions in the source
variables
\bg
Z \hs = \hs \sum_n \; a_n \, \phi_n (\{J\})
\ed
where $a_n$ are the unknowns of the problem.
A particularly simple choice is polynomial functions.
It should be emphasized that other choices
are preferred for more complicated systems.
We consider this possibility in Section Seven.

The number of independent unknown coefficients
can be reduced
by exploiting the symmetries of the lattice.
For polynomial expansions,
$Z$ can be constructed to be invariant under
the symmetry group of the lattice
\bg
Z \hs = \hs \sum_{n,m} \; a_{n,m} \, P_{n,m}
\ed
where  $P_{n,m}$  are invariant polynomials
in the source variables of order $n$
and with $m$ invariant classes for a given order.  The
coefficients $a_{n,m}$ are to be determined.  The number of
invariant classes for a given order depends on both the number
of lattice sites and on the number of symmetry operations for a
given lattice.  For example, higher dimensional lattices have
larger symmetry groups and therefore have fewer independent unknowns.
These polynomials
form a lattice invariant basis upon which we can construct
solutions for $Z$.

In order to solve the differential equations,
we must specify boundary conditions.
Normalization
of the vacuum amplitude implies
\bg
Z[J=0]  = 1
\ed
and we exclude symmetry breaking
\bg
\frac{dZ}{dJ(i)}  |_{J=0} =  \langle \phi(i) \rangle = 0.
\ed
We cannot specify the second derivatives of $Z$ because
they correspond to the two-point functions.
Instead, w truncate $Z$ at some finite order $M$.
The truncation order is dictated by
computational constraints.
As a boundary condition, truncation guarantees that as $g\ra0$ in
the interaction theory, the solution reduces to the free field.
This is
possible only for path integral formulations constrained
to real integration contours \cite{garcia,garciaguralnik,cooper}.
Truncation in this sense not
only sets a boundary condition, but also introduces an
approximation scheme.  This scheme
is made systematic by truncating at successively higher orders, and then
taking the limit as $M$ goes to infinity.

We fit our trial solution to the differential equations
using the Galerkin method.
The truncated polynomial is an approximate solution to
the differential equations where the error due to truncation
is called the residual $R$ \cite{fletcher}.
In order to minimize $R$, we define an inner product in the source space
\begin{equation}
(g,f) = \int_{-\epsilon}^{\epsilon}
g(J(1) \ldots J(N))f(J(1) \ldots J(N))  \, [dJ]
\end{equation}
where the integration is over all $J(i)$, and $\epsilon$ is considered small.
Requiring the inner product of $R$ with linearly independent
test functions $T_{k}$ to vanish
\begin{eqnarray}
(R,T_{1}) & = & 0 \nonumber \\
\vdots & &   \\
(R,T_{j}) & = & 0 \nonumber
\end{eqnarray}
generates a set of linear algebraic equations.
We can
generate as many of these equations as we like as long as the
$T_{k}$ are linearly independent.

In analogy with a variational principle,
we chose the test functions as
\bg
T_{k} \hs = \hs \frac{d}{dJ(1)} \; P_{n,m}
\ed
where the $P_{n,m}$ are the invariant polynomial that formed the basis
for $Z_{T}$ .  This choice guarantees that we always have as many
equations as unknowns, and experience has shown that it gives rapid
convergence of $Z_{T}$ to the exact solution.
We construct as many equations
as there are independent unknowns
in our power series solution
These equations can be solved by a single matrix inversion where
the resulting coefficients are the lattice Green's functions.

\section{Fermion Formulation}

We now construct the lattice functional equations for fermions.
As with the bosonic case, we appeal to a generating function $Z$
to calculate the consequences of our fermionic field theory.
We begin by considering a $1D$ system of spinless
free fermions in the presence of external sources.  We will see
that the inclusion of interactions requires only minor modifications.
Our system has dynamics described by the following action
\bg
S = \sum_{i}^{N} \,
\frac{1}{2} \bar{\psi}(i)  [  \psi(i + 1) - \psi(i - 1) ]
+  M \bar{\psi}(i) \psi(i)
+  [ \, \be(i) \psi(i) + \bar{\psi}(i) \e(i) \, ].
\ed
where the sum runs over an $N$ site lattice and
periodic boundary conditions are assumed.
Grassmann sources $\be(i),\e(i)$ live on each site of the lattice
and are coupled to the fields.  Taking the variation of the action
with respect to the fields, we obtain the EOM for the field
operators $\psi(i)$
\bg
\frac{1}{2} [ \psi(i+1) - \psi(i-1) ] + M \psi(i) + \e(i) = 0.
\ed
There are analogous equations for the conjugates. Taking vacuum
matrix elements of this equation, and identifying
expectation values of fields with Grassmann Source derivatives
of the generating functional $Z[\be,\e]$.
\bg
\frac{\partial Z}{\partial \e(i)} =  \langle \bar{\psi}(n) \rangle \; \;
\; \; \; \; \; \;\frac{\partial Z}{\partial \be(n)} =  - \, \langle \psi(n)
\rangle,
\ed
gives the lattice equations for $Z$.
\bg
\frac{1}{2} [ \frac{\partial Z}{\partial \be(i+1)}
            - \frac{\partial Z}{\partial \be(i-1)} ]
+ M \frac{\partial Z}{\partial \be(i)} - \e(i) \, Z = 0.
\ed
As with the boson case, there is one equation (and its conjugate)
for each site of the lattice.  Our fermion field theory is completely
specified by this set of coupled Grassmann PDEs for $Z$
plus boundary conditions. The
Green's functions can be calculated as derivatives of $Z[\be,\e]$ about
$\be,\e=0$.

We solve these equations with a Grassmann power series
\bg
Z = \sum_{n_{i},\bar{n}_{i} = 0,1} a_{\bar{n}_{1},n_{1}
       \cdots \bar{n}_{N},n_{N}}
       \be(1)^{n_{1}} \cdots \be(N)^{n_{N}} \,
       \e(1)^{n_{1}}  \cdots \e(N)^{n_{N}}.
\ed
Because of Fermi statistics, $Z$ is at most linear in each source
variable.  This means that for finite systems, the power series
terminates at a finite order independent
of any approximation such as truncation.  Therefore, exact solutions
exist for finite series.
To reduce the number of independent
coefficients in this series, we take
advantage of symmetries of the system.  In addition to the
space-time symmetries, the Action conserves fermion number
\bg
\psi(i) \ra e^{i \theta} \psi(i) \no
\ed
and is invariant under charge conjugation
\bg
\psi(i) \ra - \bar{\psi}(i)  \; \; \; \; \; \bar{\psi}(i) \ra \psi(i). \no
\ed
For bipartite lattices, we have the additional symmetries
\bg
\psi(i) \ra - \psi(i) \; \; \; \;  {\rm if \; i=odd}    \no  \\
\psi(i) \ra   \psi(i) \; \; \; \;  {\rm if \; i=even}.  \no
\ed
We construct the terms in our series explicitly to reflect
all these symmetries.

\section{Exact Solutions for Four Site Model}

To be specific, consider a free fermion field on four sites.
We place complex spinless fields and
sources on each site.
This gives eight coupled Grassmann differential equations for $Z$
including the conjugate equations.
Since they are all related by symmetry operations, it is sufficient
to use only one equation to constrain our solution
as long as $Z$ is constructed explicitly to
reflects these symmetries.
We choose
\bg
\frac{1}{2} [ \frac{\partial Z}{\partial \be(2)}
            - \frac{\partial Z}{\partial \be(4)} ]
+ M \frac{\partial Z}{\partial \be(1)} - \e(1) \, Z = 0.
\ed
With four complex sources in our system, there are
eight independent degrees of freedom.  Since
$Z$ can be at most linear in each Grassmann variable, the polynomial
expansion can be at most eighth order.  All higher terms vanish due
to statistics.  Therefore for four sites, we can find an {\em exact} solution
with only an eighth order Grassmann series.  This polynomial has
69 terms which can be grouped into 14 independent symmetry classes.
For example, at second order we have
\bg
Z  = 1 + a_1 \, [ \be(1) \e(1) + \be(2) \e(2)
                + \be(3) \e(3) + \be(4) \e(4)] \no \\
 + a_2 \, [ \be(2) \e(1) + \be(3) \e(2) + \be(4) \e(3) + \be(1) \e(4)  \no \\
          - \be(1) \e(2) - \be(2) \e(3) - \be(3) \e(4) - \be(4) \e(1) ] \no \\
 + a_3 \, [ \be(3) \e(1) + \be(4) \e(2)
          + \be(1) \e(3) + \be(2) \e(4) ]
\ed
where the coefficients are the Green's functions
\bg
a_1 & = & \langle \bar{\psi}(1) \psi(1) \rangle  \no \\
a_2 & = & \langle \bar{\psi}(2) \psi(1) \rangle   \\
a_3 & = & \langle \bar{\psi}(3) \psi(1) \rangle  \no
\ed
Since this is an exact solution, no Galerkin procedure is needed.
We can apply the Grassmann equation to the full polynomial, set
all the coefficients of the resulting power series to zero,
and solve the resulting system of equations.  For the exact
solution, we have as many independent equations as unknowns
and our set of algebraic equations for the coefficients of $Z$
is completely specified.
The result for the free two-point function is
\bg
a_1 &  = & \frac{2 M^2 + 1}{2 M (M^2 + 1)} \no \\
a_2 &  = & \frac{1}{2 (M^2 + 1)}  \\
a_3 &  = & \frac{1}{2 M (M^2 + 1)} \no
\ed
The complete results for all the Green's functions can
be found in the Appendix.  These results agree identically
with these from inverting the propagator matrix, or equivalently from
the lattice Fourier transformed solution to the Green's
function equation.

We can generalize our action to include self-interactions
\bg
S & = & \sum_{i}^{N} \,
\frac{1}{2} \bar{\psi}(i)  [  \psi(i + 1) - \psi(i - 1) ]
 +  M \bar{\psi}(i) \psi(i)   \no \\
&& +  \frac{G}{2} \bar{\psi}(i) \psi(i) [ \bar{\psi}(i+1) \psi(i+1)
                                     + \bar{\psi}(i-1) \psi(i-1) ]  \no \\
&& +  [ \, \be(i) \psi(i) + \bar{\psi}(i) \e(i) \, ].
\ed
In terms of the lattice functional equations, this means
adding third order derivative terms to the differential
equations
\bg
- \frac{G}{2} \, \frac{\delta^3 \, Z}{\delta \be(i)
                                   \delta \e(i+1) \delta \be(i+1)}
- \frac{G}{2} \, \frac{\delta^3 \, Z}{\delta \be(i)
                                   \delta \e(i-1) \delta \be(i-1)}
\ed
We solve the interacting model just as we did for the free case.
Again, we find an exact solution yielding
the following coefficients
\bg
a_1 &  = & \frac{2 M^3 + (2 G + 1) M}{2 M^4 + (4 G + 2) M^2 + G^2 + G} \no \\
a_2 &  = & \frac{2 M^2 + G}{4 M^4 + (8 G + 4) M^2 + 2 G^2 + 2 G}  \\
a_3 &  = & \frac{M}{2 M^4 + (4 G + 2) M^2 + G^2 + G} \no
\ed
Notice that the coefficients respect the free field limit as
$g \ra 0$, and therefore correspond to the path integral solution.

\section{Galerkin Method for Fermions}

For larger lattices, even though the power series is
self-terminating, it becomes too unwieldy for practical calculations.
Therefore, we truncate the exact expansion
at some lower order, and determine the coefficients numerically.
As in the boson case, we have the problem of inconsistent equations.
While the series expansion for $Z$ contains the full symmetry group of the
Action, the individual Grassmann differential operators are invariant
under only a subgroup.  The algebraic relations for the
coefficients of the truncated $Z$ are therefore overdetermined.
This implies that an approximate procedure is needed
to determine the coefficients.

The Galerkin procedure is based on an attempt
to minimize the error function, or the residual $R$, about
the origin in source space
\bg
R = \hat{D}^{F}_{i} \,Z_{T} [\be,\e]
\ed
where $\hat{D}^{F}_{i}$ is the fermionic source operator centered
at site $i$ and $Z_{T} [\be,\e]$ is the truncated partition function
that depends on the Grassmann sources $\be,\e$.  The residual
represents the error induced by truncation.
For bosons, in order to minimize the error,
we defined an inner product in the space spanned by
functions of the source variables $\{J(1), \ldots, J(N)\}$.
\bg
(\,g,f\,) \hs = \hs \int_{-\epsilon}^{\epsilon} \ldots
\int_{-\epsilon}^{\epsilon}
g(J(1) \ldots J(N)) \; f(J(1) \ldots J(N)) \;
dJ(1) \ldots dJ(N)
\ed
We demand that inner products of $R$ with linearly
independent test functions $\{T_j\}$ vanish, and solved the
resulting algebraic equations.  In this way, we have fitted
our approximation to a solution of the differential equation.
Solutions obtained in the manner are called ``weak solutions"
where convergence is guaranteed in the mean.

As a first attempt
to mimic this procedure for fermions, we
could begin by defining an inner product in the Grassmann
function space \cite{berezin}.
Unfortunately, there are several problems with this approach.
Principally, taking a Grassmann inner product
of $R$ with Grassmann test functions
is a very severe operation.
In fact, the Grassmann inner product
kills off most of the equations that we want to solve.
Furthermore, recall that for bosons, the Galerkin
method gave average solutions in an $\epsilon$-neighborhood of
the origin; they represented ``weak" solutions to the differential equations.
Since Grassmann integration is really a formal construct with no
measure theoretic interpretation, then in what sense
are they average solutions?

Instead, we propose to sidestep all these difficulties by making
a few simple observations.  First of all, the Grassmann sources
which create all of the difficulties carry virtually no physics.
All the information about a QFT is contained in the coefficients
of $Z$, i.e. the Green's functions,
which are simple real numbers.  The sources are arbitrary,
and in fact, for fermionic problems, their only real role is to
give the correct signs in the Green's functions
equations for the coefficients.
Once we have obtained
$R$, all the physical information has been completely specified.
Any manipulations of $R$ which are mathematically consistent
does not alter the physics.

With these observations in mind, we
propose the following alternative Galerkin method for fermions.
Once we have $R$, we relax the anticommutivity of all source variables,
and allow them to be integrated in the usual way, using the boson
inner product.  Test functions are constructed as
derivatives of scalar polynomials that have the same symmetry
as the fermion invariant polynomials.  The rest of the calculation
proceeds identically as for the boson case; inner products of
$R$ with scalar test functions gives algebraic relations for the
coefficients of $Z$.  After solving this system of equations,
$\epsilon$ is taken to zero.

A more compact way to view this procedure is to consider
each source variable as being decomposed into a scalar
piece times a Grassmann unit vector
\bg
\e(i) = \mu \, \hat{\eta}(i).
\ed
The unit vector $\hat{\eta}(i)$ carries the anticommutivity and
enforces the commutation relations while $\mu$ acts like a length
that can be integrated in the usual sense.  In this framework,
the functional
equations act on the Grassmann piece while the Galerkin inner
product operates on the scalar component.

\section{Numerical Solutions}

We test our truncation/Galerkin procedure against the
four site model discussed previously.
Since we can solve this system exactly,
it is useful for checking the
convergence of our method.
Solutions for this model were found using an eighth order
Grassmann polynomial.  To examine the convergence of our method,
we first truncate this series at fourth order.  Since this is
not an exact solution, we use a Galerkin approach to find an
approximate one.  The expressions
for the coefficients are complicated rational functions of the parameters.
Setting $M=1.0$, the results are
\bg
a_1 & = & \frac{2117446 \, G + 7294107}{1786968 \, G^2
                          + 9742031 \, G + 9725476} \no \\
a_2 & = & \frac{3092325 \, G + 4862738}{2 (1786968 \, G^2
                          + 9742031 \, G + 9725476)}\\
a_3 & = & \frac{330478 \, G + 2431369}{1786968 \, G^2
                          + 9742031 \, G + 9725476} \no
\ed
These should be compared against the exact solution in Eq (6.78).
Similarly, we truncate at sixth order, finding
\bg
a_1 & = & \frac{3 \,(618904 \, G + 510825)}
               {1345887 \, G^2 + 3667541 \, G + 2043300} \no \\
a_2 & = & \frac{(G + 2) \, (278354 \, G + 510825)}
               {2 \, (1345887 \, G^2 + 3667541 \, G + 2043300)}  \\
a_3 & = & \frac{278354 \, G + 510825}
               {1345887 \, G^2 + 3667541 \, G + 2043300} \no
\ed
To see exactly how these answers converge, we substitute a range
of values for $G$ and examine the behavior of the coefficients.
Results for the two-point functions
are displayed in the tables.

\begin{table}
\begin{center}
\begin{tabular}{|cccc|} \hline
   g     & 4th Order  & 6th Order  &  Exact   \\ \hline

   0     & 0.75       & 0.75       & 0.75     \\ \hline
   0.1   & 0.7003     & 0.7089     & 0.7095   \\ \hline
   0.5   & 0.5553     & 0.5840     & 0.5926   \\ \hline
   1.0   & 0.4428     & 0.4803     & 0.5      \\ \hline
  10     & 0.0996     & 0.1160     & 0.1494   \\ \hline
\end{tabular}
\end{center}

\caption{Convergence of fermionic Galerkin method
for $\langle \bar{\psi}(1) \psi(1) \rangle$}
\end{table}

\begin{table}
\begin{center}
\begin{tabular}{|cccc|} \hline
   g     & 4th Order  & 6th Order  &  Exact  \\ \hline

   0     & 0.25       & 0.25       & 0.25    \\ \hline
   0.1   & 0.2413     & 0.2334    & 0.2328   \\ \hline
   0.5   & 0.2130     & 0.1928    & 0.1852   \\ \hline
   1.0   & 0.1871     & 0.1677    & 0.15     \\ \hline
  10     & 0.0626     & 0.1141    & 0.0390   \\ \hline
\end{tabular}
\end{center}

\caption{Convergence of fermionic Galerkin method
for $\langle \bar{\psi}(2) \psi(1) \rangle$}
\end{table}

\begin{table}
\begin{center}
\begin{tabular}{|cccc|} \hline
   g     & 4th Order  & 6th Order  &  Exact  \\ \hline

   0     & 0.25       & 0.25       & 0.25     \\ \hline
   0.1   & 0.2299     & 0.2223     & 0.2217   \\ \hline
   0.5   & 0.1726     & 0.1543     & 0.1481   \\ \hline
   1.0   & 0.1299     & 0.1118     & 0.1      \\ \hline
  10     & 0.0201     & 0.0190     & 0.0065   \\ \hline
\end{tabular}
\end{center}

\caption{Convergence of fermionic Galerkin method
for $\langle \bar{\psi}(3) \psi(1) \rangle$}
\end{table}

For large lattices, we need to perform calculations numerically.
We calculated the interacting propagator for an eight site
lattice in one dimension.  Results are presented for a Grassmann
polynomial truncated at fourth and sixth order.
These are nontrivial calculations since
the fourth order polynomial alone has 124 coefficients.
These coefficients must be determined by the Galerkin method.

The results are checked against a mean field calculation.
In general, mean field theory is not exact, but
represents only a single pole
approximation to the propagator.
Our propagator, on the other hand, should
include contributions from all poles.   Thus, only loose
comparisons should be made between the mean field results
and the Source Galerkin numbers.

It should be noted that
calculations with the Source Galerkin method are very
efficient, especially when compared to Monte Carlo.
Typically, for lattice gauge theory, the fermions
must be quenched to make calculations tractable.
Here, determination of the interacting propagator
for a system of dynamical fermions presents
no special difficulty.
The bulk of a calculation involves
a single matrix inversion for a given set of parameters.
This is
in contrast to Monte Carlo where many sweeps through the lattice
are necessary to reduce statistical error.  As the plots show,
the Source Galerkin calculations are very clean.
They show rapid convergence compared to the mean
field results even at
intermediate couplings and using only low order polynomials.

\begin{table}
\begin{center}
\begin{tabular}{|ccc|} \hline
 4th Order & 6th Order  &  Mean Field  \\ \hline

 0.5458    & 0.5831     & 0.6287       \\ \hline
 0.2288    & 0.2104     & 0.2218       \\ \hline
 0.0883    & 0.0753     & 0.0797       \\ \hline
 0.0177    & 0.0329     & 0.0243       \\ \hline
 0.0166    & 0.0267     & 0.0196       \\ \hline
\end{tabular}
\end{center}

\caption{Interacting Propagator on 8 Sites with M=1.0, g=0.5}
\end{table}

\clearpage

\begin{figure}
\centerline{\psfig{figure=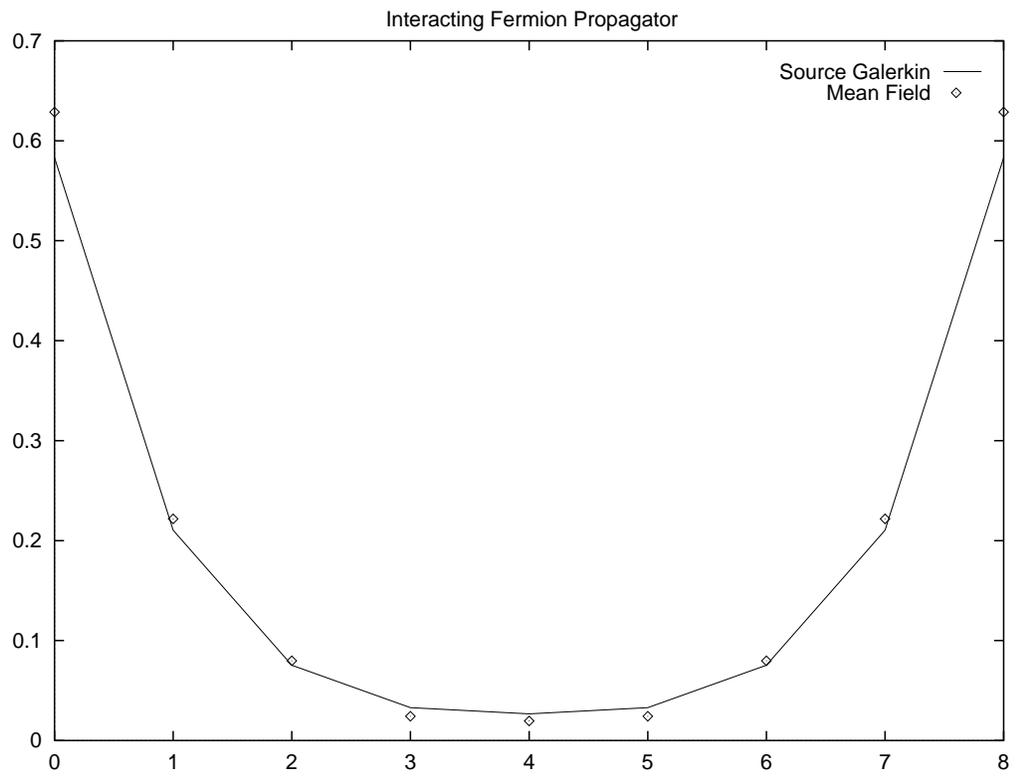,height=4in}}
\caption{8 site lattice with M=1.0 and G=0.5}
\end{figure}

\clearpage

\section{Leading Behavior of the Gross-Neveu Model}

Since the computational complexity grows rapidly with the
number of sites, we need to go beyond simple polynomial
expansions to examine larger systems.
It is important to note that we are not
restricted to power series expansion.  Any complete
set of functions is allowed where the choice is restricted only
by convenience.
We exploit this flexibility with
respect to the choice of expansion functions.
A more general choice is
a Gaussian ansatz.
This choice leaves only one unknown parameter to
solve, independent of the lattice size.
We will use this ansatz to
examine the structure of a nontrivial model of interacting
relativistic fermions.

The Gross-Neveu model \cite{gross} is a (1+1)-dimensional renormalizable
field theory of N species of Dirac fermions interacting through
a 4-fermion term.  The action is given by
\bg
L = \sum_{j=1}^{N} \, [ \bar{\psi^{(j)}} \slashD \psi^{(j)}
- \frac{1}{2} g^2 \, (\bar{\psi^{(j)}}  \psi^{(j)})^2 ]
\ed
where the four-fermion term is often replaced by a quadratic term with the
help of a real scalar field $\sigma$
\bg
L = \sum_{j=1}^{N} \, [ \bar{\psi^{(j)}} \slashD \psi^{(j)}
+ \frac{1}{2 g^2} \, \sigma^2
+ \sigma \, \bar{\psi^{(j)}} \psi^{(j)} ].
\ed
The action displays
a discrete chiral symmetry
\bg
\psi & \ra & \gamma_5 \psi  \no \\
\bar{\psi} & \ra & - \bar{\psi} \gamma_5 \no \\
\sigma & \ra & - \sigma \no
\ed
where $\gamma_5 = i \, \gamma_1 \, \gamma_2$.  The model can be studied
analytically by the saddle point method, leading to a $1/N$
expansion for $g^2 \, N$ fixed.  At leading order, the
chiral condensate is given by
\bg
\langle \, \sigma \, \rangle = \Lambda \, e^{-\pi/(g^2 \, N)}.
\ed
The theory describes a free fermion
where the dynamical mass $m_{f} = \langle \sigma \rangle$
results from spontaneous
chiral symmetry breaking.

We apply
the Source Galerkin method to this model.
For simplicity, we employ staggered fermions.  To
probe the details of the system, we need lattices considerably
larger than the ones used in the $1D$ example.  For such lattices,
construction and manipulation of the Grassmann invariant polynomials
becomes difficult. Instead, we take advantage of the extreme flexibility
of the Galerkin method with regards to the choice of expansion functions.
Following a similar calculation for scalar
$\phi^4$ \cite{garcia}, we
determine the leading asymptotic behavior for vanishing sources
$J \ra 0$ of the GNM.

After including sources $(\be_{n},\e_{n},S_{n})$
for the fermion and auxiliary fields, we vary the staggered GN action
to obtain the equations of motion
\bg
(D_{n,m} + \sigma_{n}) \, \chi^{\alpha}_{n} & = & \e^{\alpha}_{n} \no \\
\frac{\sigma_{n}}{2 \lambda} + \sum_{n=1}^{\alpha} \,
\bar{\chi}_{n}^{\alpha} \chi_{n}^{\alpha}  & = & S_{n}
\ed
where
\bg
D_{n,m} = \frac{1}{2} \, [ \delta_{n,m+\hat{1}} - \delta_{n,m-\hat{1}} ]
 + \frac{1}{2} \, (-)^{n_1}\, [ \delta_{n,m+\hat{2}} - \delta_{n,m-\hat{2}} ]
\ed
is the Susskind differential operator with $\hat{1},\hat{2}$ being unit
vectors along the two Euclidean directions
and where $\lambda = g^2 N$.
After taking matrix elements and inserting functional definitions, we
obtain the lattice functional equations
\bg
D_{n,m} \, \frac{\delta Z}{\delta \be^{\alpha}_{n}}
+ \frac{\delta^2 Z}{\delta S_{n} \delta \be_{n}^{\alpha}} &
 = & \e_{n}^{\alpha} Z  \no \\
\frac{1}{2 \lambda} \frac{\delta Z}{\delta S_{n}}
+ \sum_{n=1}^{\alpha} \,
\frac{\delta^2 Z}{\delta \be_{n}^{\alpha} \delta \e_{n}^{\alpha}}
& = & S_{n} Z
\ed
We solve these equations with a Gaussian in the sources
\bg
Z = e^{W(\be,\e)}
\ed
and find the following equations for $W(\be,\e)$
\bg
[ D_{n,m} + \frac{\delta W}{\delta S_{n}} ] \,
\frac{\delta Z}{\delta \be^{\alpha}_{n}}  =  \e_{n}^{\alpha} \no \\
\frac{N}{2 \lambda} \frac{\delta W}{\delta S^{n}}
+ \sum_{n=1}^{\alpha} \, [
\frac{\delta^2 W}{\delta \be_{n}^{\alpha} \delta \e_{n}^{\alpha}} +
(\frac{\delta W}{\delta \be_{n}^{\alpha}})
(\frac{\delta W}{\delta \e_{n}^{\alpha}}) ]  =  S_{n}.
\ed
Expanding $W$ in a 2nd order power series in the Grassmann sources gives
\bg
W(\be,\e) = \be_{k} G(k - j) \e_{j}
\ed
where $G(k - j)$ is the interacting propagator.  Since $\sigma_{n}$ is
not dynamical, we ignore it at this order, and
set $\delta W/ \delta S$ which measures chiral symmetry
breaking equal to a constant
\bg
\frac{\delta W}{\delta S} = \langle \, \sigma \, \rangle = \sigma_0.
\ed
Keeping only linear terms in the sources, we obtain a dynamical equation
\bg
[ D_{n,m} + \sigma_0 ] \, G(n-m) = \delta_{n,m}
\ed
as well as a constraint equation.
\bg
\frac{N}{\lambda} = G(0).
\ed
The dynamical equation suggests that $G(n - m)$ has the form of
a free staggered propagator with mass $\sigma_0$.  We can solve
this equation by Fourier transform
\bg
G(n-m) = \frac{1}{L^2} \, \sum_{p_1=0}^{L-1} \sum_{p_2=0}^{L-1} \,
\frac{\sigma_0 + i \sin(p_1)
+ i (-)^{m_1} \sin(p_2)}{\sigma_0^2 + \sin^2(p_1) + \sin^2(p_2)} \,
e^{ip (n-m)}
\ed
where $n,m,p$ are two dimensional vectors.
Putting this expression into the constraint gives a gap equation
for $\sigma_0$ that must be solved self-consistently
\bg
\frac{1}{\lambda} = \frac{1}{L^2} \, \sum_{p_1=0}^{L-1} \sum_{p_2=0}^{L-1} \,
\frac{1}{\sigma_0^2 + \sin^2(p_1) + \sin^2(p_2)}.
\ed
This type of equation is familiar from large $N$ expansions.


It is important to note that this solution
has been obtained by fitting a Gaussian
\bg
Z = \exp \omega(\be,\e)
\ed
to the functional equations
where we kept only second order terms in the expansion for $W$
\bg
\omega(\be,\e) = \be_{j} G_{j,k} \e_{k} + \ldots
\ed
At this level of approximation, the procedure
may seem rather trivial.
The results we have obtained are closely related
to the usual large N expansion results.
However, it should be noted that
our numerical procedure has determined the
leading order solution of a nontrivial
fermionic problem.
Systematic improvements of approximations are then possible
where the Gaussian is only the leading contribution.
By considering $Z$ as a Gaussian multiplied by a power series,
it is possible to improve our leading order result.
The Galerkin method can be used to
determine the coefficients of the polynomials.

\section{Conclusion}

We have presented a new numerical method to solve
lattice fermion theories.
It is based on the differential formulation of QFT in the
presence of an external source.
By examining the functional differential equations for a theory
on a finite lattice, we obtained a set of coupled
Grassmann PDEs for the
generating functional $Z$.  For nonlinear field theories with
polynomial interactions, these equations are always linear.
Once $Z$ is obtained, the lattice Green's
functions can be extracted by functional differentiation.

To construct solutions, we can expand $Z$
on {\em any} complete set of functions
in the source variables $\{\eta_i\}$.  A particularly simple choice is
polynomial functions.  We saw that these functions gave very
rapid convergence even using low order polynomials.
Calculations were efficient to perform and produced very clean
numbers.
The bulk of any calculation involved only a single matrix inversion.
The matrix to be inverted becomes large,
but unlike Monte Carlo, there is only one inversion for a given
set of parameters.

Due to computational complexity, polynomial basis
functions are limited to small lattices.
For larger systems,
more general schemes are possible.
The Galerkin procedure of fitting approximate solutions to the
functional equations allows tremendous freedom in choosing
expansion functions.
We illustrated one possibility by
using a Gaussian to extract the leading behavior
of the solution.
In this formulation, we had only one unknown parameter,
independent of the size of the lattice.
Systematic corrections can then be calculated.
We will have more to say about other generalized approaches
in future communications.

This method is deterministic, and therefore,
does not suffer from many of the problems usually associated
with numerical solutions of fermionic problems.
For example,
Monte Carlo simulations
of lattice gauge theories have difficulty including dynamical
fermions.
But with the Source Galerkin method, calculation of the
interacting propagator presents no special complication.
In addition,
fermionic systems at nonzero chemical potential have problems
defining a positive definite probability measure for simulations
due to the ``minus sign" problem.
Since our method is deterministic, this is not an issue.

\clearpage
\section{Appendix A: Exact Coefficients for Four Site Model}
\bg
A1 & = & \frac{2 \, M^2  + (2 \, G + 1) \, M}{2 \, M^4 + (4 \, G
 + 2) \, M^2 + G^2 + G} \no \\ \no \\
A2 & = & \frac{2 \, M^2  G}{2 \, M^4 + (4 \, G
 + 2) \, M^2 + G^2 + G} \no \\ \no \\
A3 & = & \frac{M}{2 \, M^4 + (4 \, G
 + 2) \, M^2 + G^2 + G} \no \\ \no \\
A4 & = & \frac{4 \, M^2 + 2 \, G + 1}{4 \, M^4 + (8 \, G
 + 4) \, M^2 + 2 \, G^2 + 2 \, G} \no \\ \no \\
A5 & = & \frac{M}{2 \, M^4 + (4 \, G
 + 2) \, M^2 + G^2 + G} \no \\ \no \\
A6 & = & \frac{1}{4 \, M^4 + (8 \, G
 + 4) \, M^2 + 2 \, G^2 + 2 \, G} \no \\ \no \\
A7 & = & \frac{M}{2 \, M^4 + (4 \, G
 + 2) \, M^2 + G^2 + G} \no \\ \no \\
A8 & = & \frac{1}{4 \, M^4 + (8 \, G
 + 4) \, M^2 + 2 \, G^2 + 2 \, G} \no \\ \no \\
A9 & = & \frac{2 \, M^2}{2 \, M^4 + (4 \, G
 + 2) \, M^2 + G^2 + G} \no \\ \no \\
A10 & = & 0 \no \\ \no \\
A11 & = & \frac{2 \, M}{2 \, M^4 + (4 \, G
 + 2) \, M^2 + G^2 + G} \no \\ \no \\
A12 & = & \frac{1}{2 \, M^4 + (4 \, G
 + 2) \, M^2 + G^2 + G} \no \\ \no \\
A13 & = & 0 \no \\ \no \\
A14 & = & \frac{2}{2 \, M^4 + (4 \, G
 + 2) \, M^2 + G^2 + G} \no
\ed

\clearpage
\section{Appendix B: Fourth Order Coefficients with M=1.0}
\bg
A1 & = & \frac{2117446 \, G + 7294107}{1786968 \, G^2
        + 9742031 \, G + 9725476} \no \\ \no \\
A2 & = & \frac{3092325 \, G + 4862738}{ 2 (1786968 \, G^2
        + 9742031 \, G + 9725476)} \no \\ \no \\
A3 & = & \frac{330478 \, G + 2431369}{1786968 \, G^2
        + 9742031 \, G + 9725476} \no \\ \no \\
A4 & = & \frac{3573936 \, G + 12156845}{2 (1786968 \, G^2
        + 9742031 \, G + 9725476)} \no \\ \no \\
A5 & = & \frac{792533 \, G + 2431369}{1786968 \, G^2
        + 9742031 \, G + 9725476} \no \\ \no \\
A6 & = & \frac{2431369}{2 (1786968 \, G^2
        + 9742031 \, G + 9725476)} \no \\ \no \\
A7 & = & \frac{792533 \, G + 2431369}{1786968 \, G^2
        + 9742031 \, G + 9725476} \no \\ \no \\
A8 & = & - \frac{736530 \, G - 2431369}{2 (1786968 \, G^2
        + 9742031 \, G + 9725476)} \no \\ \no \\
A9 & = & \frac{1061159 \, G + 4862738}{1786968 \, G^2
        + 9742031 \, G + 9725476} \no \\ \no \\
A10 & = & \frac{24642 \, G}{1786968 \, G^2
        + 9742031 \, G + 9725476} \no
\ed

\clearpage
\section{Appendix C: Sixth Order Coefficients with M=1.0}
\bg
A1 & = & \frac{3 \, (618904 \, G + 510825)}{1345887 \, G^2
        + 3667541 \, G + 2043300} \no \\ \no \\
A2 & = & \frac{(G + 2) (278354 \, G + 510825)}{2 (1345887 \, G^2
        + 3667541 \, G + 2043300)} \no \\ \no \\
A3 & = & \frac{278354 \, G + 510825}{1345887 \, G^2
        + 3667541 \, G + 2043300} \no \\ \no \\
A4 & = & \frac{5 (482684 \, G + 510825)}{2 (1345887 \, G^2
        + 3667541 \, G + 2043300)} \no \\ \no \\
A5 & = & \frac{278354 \, G + 510825}{1345887 \, G^2
        + 3667541 \, G + 2043300} \no \\ \no \\
A6 & = & \frac{278354 \, G + 510825}{2 (1345887 \, G^2
        + 3667541 \, G + 2043300)} \no \\ \no \\
A7 & = & \frac{278354 \, G + 510825}{1345887 \, G^2
        + 3667541 \, G + 2043300} \no \\ \no \\
A8 & = & \frac{278354 \, G + 510825}{2 (1345887 \, G^2
        + 3667541 \, G + 2043300)} \no \\ \no \\
A9 & = & \frac{2 (278354 \, G + 510825)}{1345887 \, G^2
        + 3667541 \, G + 2043300} \no \\ \no \\
A10 & = & 0
         \no \\ \no \\
A11 & = & \frac{1021650}{1345887 \, G^2
        + 3667541 \, G + 2043300} \no \\ \no \\
A12 & = & \frac{510825)}{1345887 \, G^2
        + 3667541 \, G + 2043300} \no \\ \no \\
A13 & = & \frac{556708 \, G}{1345887 \, G^2
        + 3667541 \, G + 2043300} \no
\ed

\section{Acknowledgements}

We would like to thanks Andrew Sonnenschein for writing
symbolic algebra routines useful for calculations
in this paper. We are indebted to Santiago Garcia for many discussions.
GSG would like to thank Vance Faber for the Hospitality of C-3 at Los
Alamos National Laboratory where some of this work was done.
We also wish to thank Brad Marston for useful discussions
during the early phase of this work.

\end{document}